\documentclass[twocolumn]{aastex701}
\shorttitle{Dust along the line of sight of GX 340+0}
\shortauthors{Rogantini et al.}
\usepackage{tabularx}
\usepackage{graphicx}
\usepackage[referable]{threeparttablex}
\usepackage{txfonts}
\usepackage{multirow}
\usepackage{amssymb}
\usepackage{units}
\usepackage{float}
\usepackage[caption = false]{subfig}
\usepackage{hyperref}
\usepackage{natbib}
\bibpunct{(}{)}{;}{a}{}{,}
\usepackage{url}
\usepackage{array}
\usepackage{xspace}
\usepackage{enumitem,booktabs}

\newcommand{\gx}{GX~340+0\xspace}
\usepackage{xspace}

\newcommand{\astrosat}{{ASTRO-SAT}\xspace}
\newcommand{\chandra}{{\it Chandra}\xspace}

\newcommand{\xmm}{XMM{\it-Newton}\xspace}
\newcommand{\xrism}{{XRISM}\xspace}

\newcommand{\NH}{\ensuremath{N_{\mathrm{H}}}\xspace}

\newcommand{\xspec}{{\textsc{Xspec}}\xspace}
\newcommand{\spex}{{\textsc{Spex}}\xspace}

\newcommand{\sii}{\ion{Si}{1}\xspace}

\newcommand{\Si}{\ion{S}{1}\xspace}
\newcommand{\Sii}{\ion{S}{2}\xspace}
\newcommand{\Siii}{\ion{S}{3}\xspace}

\newcommand{\cstat}{\textit{C}stat\xspace}

\begin{document}

\title{Silicon, sulfur and iron in the interstellar medium: a high-resolution X-ray spectral study of \gx}

\correspondingauthor{Daniele Rogantini}
\email{danieler@uchicago.edu}

\author[0000-0002-5359-9497]{Daniele Rogantini}
\affiliation{Department of Astronomy and Astrophysics, University of Chicago, 5640 S Ellis Ave, Chicago, IL 60637, USA}
\affiliation{MIT Kavli Institute for Astrophysics and Space Research, Massachusetts Institute of Technology, Cambridge, MA 02139, USA}
\email[show]{danieler@uchicago.edu}

\author[0000-0002-5769-8441]{Claude R. Canizares}
\affiliation{MIT Kavli Institute for Astrophysics and Space Research, Massachusetts Institute of Technology, Cambridge, MA 02139, USA}
\email{}

\author[0000-0001-8470-749X]{Elisa Costantini}
\affiliation{SRON Netherlands Institute for Space Research, Niels Bohrweg 4, 2333 CA Leiden, The Netherlands}
\email{}

\author[0000-0001-9911-7038]{Liyi Gu}
\affiliation{SRON Netherlands Institute for Space Research, Niels Bohrweg 4, 2333 CA Leiden, The Netherlands}
\email{}

\author[0000-0002-4992-4664]{Missagh Mehdipour}
\affiliation{Department of Astronomy, University of Michigan, Ann Arbor, MI 48109, USA}
\email{}

\author[0000-0002-1049-3182]{Ioanna Psaradaki}
\affiliation{European Space Agency (ESA), European Space Research and Technology Centre (ESTEC), Keplerlaan 1, 2201 AZ Noordwijk, The Netherlands}
\email{}

\author[]{Norbert S. Schulz}
\affiliation{MIT Kavli Institute for Astrophysics and Space Research, Massachusetts Institute of Technology, Cambridge, MA 02139, USA}
\email{}

\author[0000-0002-8163-8852]{Sascha Zeegers}
\affiliation{SRON Netherlands Institute for Space Research, Niels Bohrweg 4, 2333 CA Leiden, The Netherlands}
\email{}

\begin{abstract}
High-resolution X-ray spectroscopy provides a powerful probe of the interstellar medium (ISM), giving direct access to the composition and physical state of dust grains and atomic species in dense environments. We present a study of the gas and dust along the line of sight to the bright low-mass X-ray binary \gx, which samples higher-density gases in the ISM. Using deep \chandra/HETG spectra, we characterize X-ray absorption fine structure from dust, gas absorption lines, and the optical depths of the Si, S, and Fe K-edges. By fitting these three edges simultaneously, we reduce degeneracies in the dust composition and find that amorphous olivine dominates the fractional contribution among the dust columns ($\sim$65\%), followed by metallic iron ($\sim$19\%), iron sulfides (pyrrhotite and troilite; $\sim$10\%), and fayalite ($\sim$5\%), with the remaining species contributing only a few percent in total. From the inferred stoichiometry, we estimate that $\sim$74\% of Fe is associated with silicates, $\sim$8\% with sulfides, and $\sim$18\% with metallic iron, suggesting that Fe is predominantly incorporated in iron rich silicate grains along this sightline. We detect \Sii\ absorption and infer a sulfur dust fraction of $\sim$35\%. We also detect absorption structure near the Ca and Ar K edges, highlighting the need for improved atomic photoabsorption data. The \chandra/HETG spectral resolution remains essential to disentangle dust and gas contributions at the Si and S K edges, providing a benchmark for dust characterization in high-density regions in the ISM.
\end{abstract}\section{Introduction} \label{sec:intro}

The role of dust in galactic evolution has been recognized for decades \citep{Greenberg63}. Interstellar dust traces key physical conditions (e.g., magnetic fields and gas temperature) and actively exchanges material with the gas phase of the interstellar medium (ISM). It regulates star and planet formation and supports the formation of simple and complex molecules. Despite major progress in infrared (IR) and ultraviolet (UV) astronomy, and in laboratory astrophysics, fundamental questions remain open on the composition, size distribution, and morphology of interstellar grains, in both diffuse and dense environments.

Silicon is among the best-studied dust-forming elements, identified through spectroscopy across multiple wavelength bands, including IR and X-rays. The 9.7~$\mu$m and 18~$\mu$m silicate features and the Si K-edge in X-rays provide strong constraints on silicate chemistry \citep{Kemper04,Min07,vanBreemen11,Zeegers19,Rogantini20}. Silicon is primarily hosted in amorphous silicates with olivine $\rm (Mg,Fe)_2SiO_4$ and pyroxene $\rm (Mg,Fe)SiO_3$ stoichiometries. Depletion studies \citep[e.g.][]{Jenkins09,Zhukovska18} show that silicon depletion correlates with gas density, indicating efficient grain growth in dense regions.

Interstellar iron remains a long-standing puzzle. In contrast to Mg and Si, which are efficiently produced in stellar dust-forming outflows, Fe is predominantly synthesized in Type Ia supernovae, which lack such outflows. As a result, a substantial fraction ($\sim 65\%$) of iron is injected into the ISM in gaseous form \citep{Dwek16}. Observationally, however, Fe is almost fully depleted from the gas phase in both warm and cold neutral media, implying rapid and efficient accretion onto dust grains \citep{Lee09,Dwek16,Zhukovska18}. Silicates alone appear to account for less than 40\% of the iron budget \citep{Poteet15}, motivating alternative reservoirs such as iron sulfides (FeS) and metallic nanoparticles \citep{Min07,Jones13,Hensley17}. X-ray spectroscopy instead indicates a higher fraction of Fe in silicates, up to $\sim$70\% \citep{Rogantini20,Psaradaki23}. Iron may also be incorporated as metallic inclusions within silicate grains, which could protect it from rapid erosion by interstellar shocks \citep{Zhukovska18}.

Sulfur is another key element with poorly constrained ISM chemistry \citep{Laas19}. In diffuse environments sulfur is largely found in ionized atomic form \citep{Savage96,Jenkins09}, while depletion onto grains is expected in dense regions such as molecular clouds. The presence of sulfur in icy and rocky solar-system bodies suggests an efficient transition from volatile sulfur in diffuse media to refractory sulfur in grains. Laboratory and observational work proposes that sulfur can be hosted in a variety of organic compounds \citep{Laas19} and in sulfur allotropes such as cyclo-octasulfur \citep[$\rm S_8$][]{Shingledecker20}. \citet{Ferrari24} recently identified IR signatures of sulfur rings, enabling searches in dark clouds with \emph{James Webb Space Telescope}. In addition, interstellar grains preserved in meteorites and interplanetary dust particles show sulfur incorporated as Fe-rich sulfides, especially in GEMS (glass with embedded metals and sulfides), where S/Si ratios reach 60--80\% of solar values \citep{Bradley94,Bradley96}. Consistent indications from the Rosetta \citep{Calmonte16} and Stardust \citep{Westphal14} missions support the presence of sulfur in interstellar grains even in diffuse environments, challenging scenarios where UV radiation efficiently removes sulfur from grain surfaces \citep{Keller10}.

High-resolution X-ray spectroscopy is a powerful probe of dust chemistry and grain physics \citep{Costantini22}. The X-ray band includes the K- or L-shell absorption features of the most abundant metals (C, N, O, Ne, Mg, Si, S, Fe). Near absorption edges, X-ray absorption fine structures (XAFS) encode information on the local bonding environment and can constrain grain composition, crystallinity, size, and shape \citep{Newville04}. Studies of bright Galactic X-ray binaries, combined with modern dust extinction models, have already provided strong constraints on the O K and Fe L-edges in diffuse media \citep[e.g.][]{Lee05,Costantini12,Westphal19,Psaradaki20,Corrales24}. At higher energies, the Mg and Si K-edges (1--2 keV) are particularly effective diagnostics of interstellar silicates \citep[e.g.][]{Zeegers19,Rogantini20}.

The S K-edge at 2.47~keV (5.02~\AA) and the Fe K-edge at 7.1~keV (1.745~\AA) are especially valuable for highly-absorbed sightlines and for identifying S- and Fe-bearing grain populations. These edges become prominent for column densities above $\sim 5 \times 10^{22}\ \rm cm^{-2}$ \citep[see Figure~1.3 of][]{Rogantini20phd}, where they represent primary ISM signatures detectable in X-ray spectra \citep{Rogantini18}. Before the launch of \xrism, available missions generally lacked the energy resolution needed to resolve fine structures at the Fe K-edge, which requires resolution of a few eV \citep[e.g.][]{Alderman17,Rogantini18}; in comparison, the \emph{Chandra} HETG resolution is $\sim$45 eV at these energies. However, simultaneous fitting of multiple edges with broadband dust extinction models can reduce degeneracies and mitigate instrumental limitations \citep{Rogantini20,Psaradaki23}.

This work presents the analysis of deep ($\sim$300~ks exposure) \chandra high-resolution X-ray spectra of \gx. The source is a persistent, bright neutron-star low-mass X-ray binary and \emph{Z} source located near the Galactic Plane at a distance of $11 \pm 3$ kpc \citep{Penninx93}. It is heavily obscured, with $\NH = 0.7$--$1.1 \times 10^{23}\ \rm cm^{-2}$ \citep{Miller16b,Zeegers19}, making it an ideal target to study dust in the interstellar medium through the Si, S, and Fe K-edges. Its high flux ($F_{2-10\:\rm keV}\simeq 10^{-8}\ \rm erg\ cm^{-2}\ s^{-1}$; \citealt{dAi09}) and extreme extinction allow tight constraints on dust properties in the higher-density phases of the ISM.

\gx was also observed during the \xrism\ performance-verification phase with \xrism/\textsc{Resolve} for $\sim$150~ks. With the gate valve closed, most of the science focused on energies $\gtrsim$2~keV \citep{Ludlam25}. In the Fe~K band, the \textsc{Resolve} spectrum revealed significant structure within the relativistic Fe line complex across spectral states: reflection models tailored for neutron-star illumination reproduce the broad components but leave residual narrow features at the $\sim$5\% level near 6.7 and 6.97~keV, suggesting an additional contribution from ionized plasma \citep{Ludlam25}. A complementary analysis based on detailed photoionization modeling decomposed the Fe~\textsc{xxv} He$\alpha$ profiles into narrow ($\sim$360~km~s$^{-1}$) and broad ($\sim$800~km~s$^{-1}$) components and identified a modest accretion-disk wind, exhibiting both emission and absorption features at $v\simeq 2735$~km~s$^{-1}$, together with a relativistic reflection component \citep{Chakraborty25}. In parallel, the high signal-to-noise \textsc{Resolve} spectrum around the S~K edge enabled a direct measurement of sulfur in both gas and solid form: after modeling the atomic S~\textsc{ii} absorption, residuals were consistent with absorption from Fe--S dust (e.g., troilite, pyrrhotite, or pyrite), yielding an S depletion of $40\%\pm15\%$ and an upper limit of $<25\%$ on the fraction of interstellar Fe bound in Fe--S compounds along this sightline \citep{Corrales25}.

With the Resolve gate valve closed, \xrism sensitivity is strongly reduced below $\sim$2~keV, and the Si~K edge cannot be observed. \chandra/HETG\ therefore remains the best instrument to study dust features in the silicon K band in detail, providing crucial constraints on silicate grains. In this work, we use deep \chandra/HETG\ spectra to constrain the dust properties along the line of sight to \gx, analysing the high-resolution grating data with the \spex\ fitting package \citep{Kaastra22}. Section~\ref{sec:observation} outlines the observations and data reduction, while Section~\ref{sec:results} presents the spectral analysis and the resulting constraints on the Si, S, and Fe K-edge absorption. We discuss the implications for dust composition and depletion in ISM environments in Section~\ref{sec:discussion}, before summarizing the main conclusions in Section~\ref{sec:summary}. All best-fit values are reported with $1\sigma$ uncertainties.

\section{Observations} \label{sec:observation}
\chandra/HETG has observed the bright X-ray binary \gx multiple times over the past 25 years. For this work, all \chandra/HETG observations obtained in Timed Exposure (TE) mode between August 2001 and August 2023 were retrieved and analyzed (Table~\ref{tab:log}). The most recent observations, carried out between January 2022 and August 2023, were awarded through a Guaranteed Time Observation program (proposal 23910648; PI: Rogantini) to study interstellar dust signatures at the Si, S, and Fe K-edges of \gx.

The HETG data were reduced with the \emph{Chandra Interactive Analysis of Observations} software \citep[{\sc ciao};][]{Fruscione06}, following the standard procedures described by TGCat \citep{Huenemoerder11}. The High Energy Grating (HEG) and Medium Energy Grating (MEG) spectra were extracted using a narrow region mask with a width factor of 18, optimized to preserve the spectral shape above $\sim7.5$~keV ($\sim1.7$~\AA). Although this choice reduces the background extraction area, the high source count rate makes the background negligible across the full bandpass.

Given the brightness of \gx (typical count rates $>50$~cts~s$^{-1}$), pileup is a significant concern \citep{Schulz16}. A fast continuous-clocking configuration is not suitable in this case because it distorts the edge fine structure (see the \chandra Proposer's Observatory Guide \footnote{\url{https://cxc.harvard.edu/proposer/POG/}}). To mitigate pileup, the zeroth order was positioned near the edge of the detector array in all observations (except ObsID~1921), reducing the effective frame time. This setup provides pileup-free first-order HEG spectra \citep{Yang22}. As a consequence, the analysis is restricted to the HEG $-1$ and MEG $+1$ orders, while the HEG $+1$ and MEG $-1$ orders are excluded. In addition, this configuration severely limits the usable wavelength coverage of higher grating orders, which in practice only cover the Fe K-edge. Given their lower effective area, the second and third orders were not included in the analysis. ObsID 1921 is affected by pile-up in the MEG first orders, and these spectra were therefore excluded from the analysis. By contrast, the HEG shows negligible pile-up contamination in the band considered, making the modeling of the edges and the associated dust features robust.

The first orders of the individual observations were combined using the {\tt combine\_grating\_spectra} tool. Since ISM absorption is not expected to vary on decadal timescales, combining the datasets is appropriate for studying interstellar spectral features. 
We examined each observation individually to define conservative HEG and MEG fitting bands. The spectral shape remains broadly stable, while the normalization varies by a factor of 2--3. We verified that this variability does not affect the dust-feature fits by jointly fitting the observations with tied ISM parameters. Broader emission features, such as Fe K$\alpha$ (not relevant to the goals of this paper) may be distorted in the stacked spectrum and should therefore be interpreted with caution.
The combined HEG and MEG spectra were fitted simultaneously, including a cross-calibration constant that remained close to unity, indicating relative calibration differences below $\sim3\%$. The HEG and MEG data were analyzed over $1.5$--$7.0$~\AA\ ($1.77$--$8.27$~keV) and $4$--$8$~\AA\ ($1.55$--$3.1$~keV), respectively. Owing to the strong absorption, the signal-to-noise ratio around the Mg K-edge ($\sim1.3$~keV) is too low and this region is excluded. Finally, to exploit the superior resolution of HEG \citep[see also][]{Yang22}, the MEG spectra were excluded in the immediate vicinity of the Si and S K-edge XAFS regions (4.95--5.05~\AA\ and 6.63--6.75~\AA, respectively). The spectra were optimally binned following \citet{Kaastra16}.

%-------------------------------------------------------------
%                                              Observation Log 
%-------------------------------------------------------------
%
\begin{table}
\caption{\chandra/HETG observation log. The total exposure time and the average count rate are shown at the bottom. The list of \chandra datasets used in this paper is, contained in~\dataset[DOI: 10.25574/cdc.570]{https://doi.org/10.25574/cdc.570}.}
\label{tab:log}
\centering
\begin{tabular}{ c c c c }
\hline
\noalign{\vskip 0.5mm}
ObsID & Date & Exp. Time [ks] & Count Rate [cts/s] \\
\noalign{\vskip 0.5mm}
\hline
\noalign{\vskip 0.5mm}
1921  & 2001-08-09 & 23.4 & 71.5 \\
18085 & 2016-06-27 & 24.1 & 53.4 \\
19450 & 2017-06-13 & 64.5 & 54.3 \\
20099 & 2017-06-19 & 60.6 & 62.7 \\
26109 & 2022-01-20 & 14.0 & 43.4 \\
26281 & 2022-01-20 & 14.0 & 43.4 \\
26110 & 2023-03-13 & 28.9 & 56.8 \\
26078 & 2023-05-25 & 26.0 & 46.2 \\
26107 & 2023-05-29 & 26.9 & 47.4 \\
26108 & 2023-08-28 & 19.5 & 57.8 \\
28875 & 2023-08-30 & 9.3  & 33.6 \\
\noalign{\vskip 0.5mm}
\hline
& & 311.2 & $\langle 51.9\rangle$ \\
\noalign{\vskip 0.5mm}
\hline
\end{tabular}
\end{table}

\section{Analysis and Results} \label{sec:results}

\begin{figure}[t]
\centering
\includegraphics[width=\columnwidth]{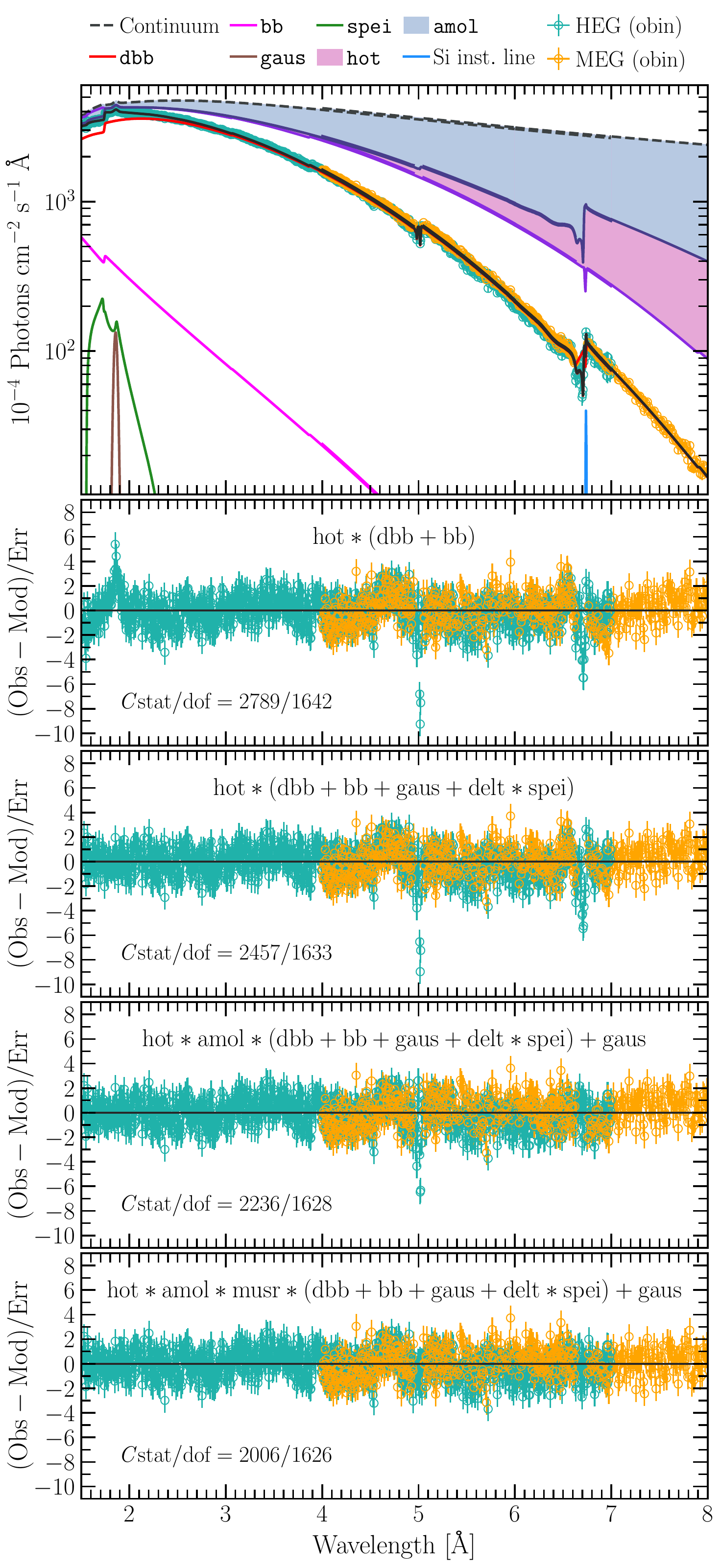}\\
\caption{\textit{Top panel:} Stacked \chandra/HETG spectrum of \gx with the best-fit model (black solid line). Individual emission components are shown as colored solid lines, while absorption components are indicated by shaded areas. The dashed black line shows the intrinsic emission before absorption. \textit{Middle and bottom panels:} Fit residuals for a sequence of models of increasing complexity, from top to bottom.}
\label{fig:continuum_residuals}
\end{figure}

\begin{figure*}[t]
\centering
\includegraphics[width=\textwidth]{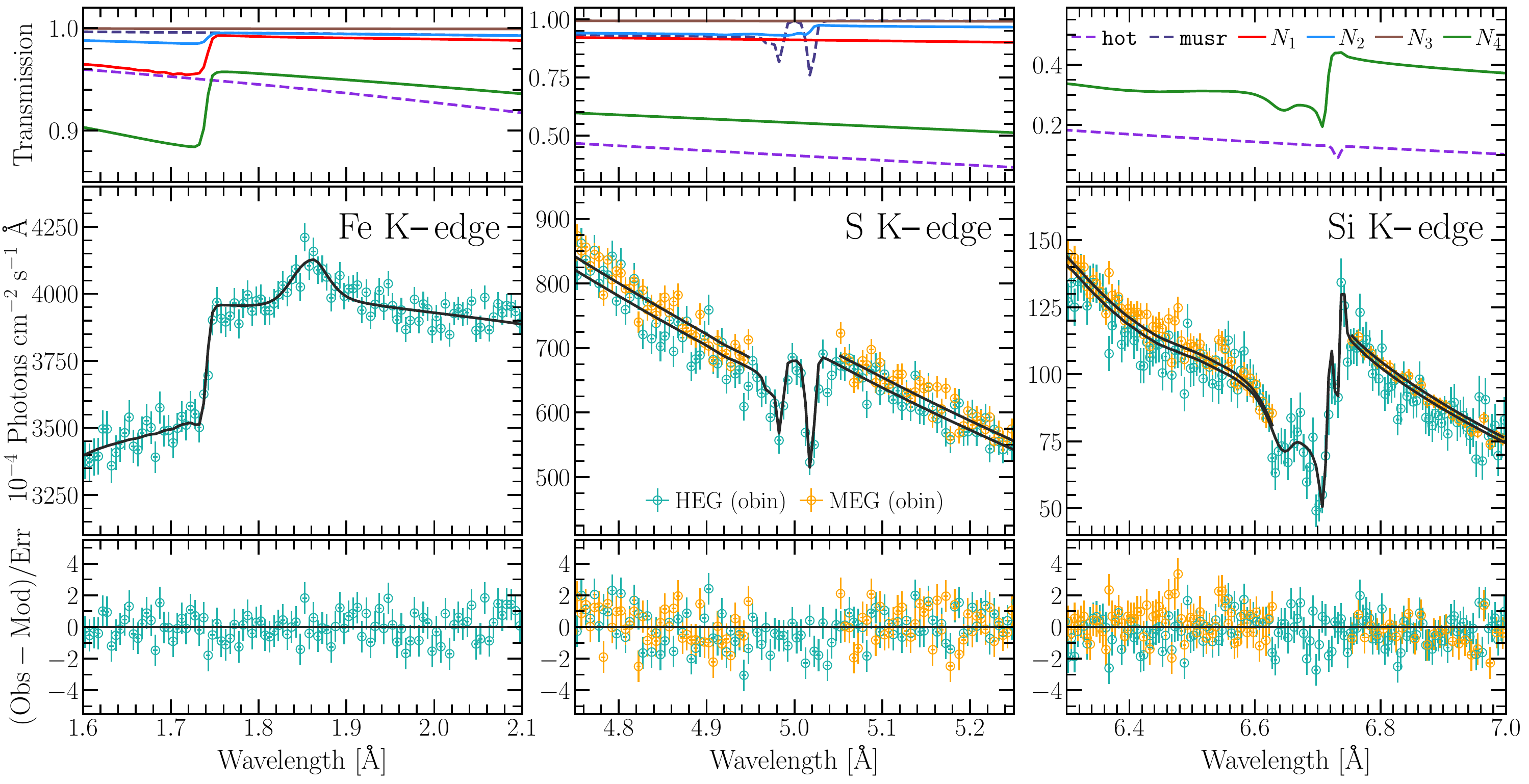}\\
\caption{Close-up view of the Fe, S, and Si K-edge regions. The stacked HETG spectra are shown in the middle panels and the residuals of the best-fit model in the bottom panels. The dominant absorption components are illustrated in the top panels as transmitted fraction (ratio of observed to incident flux).}
\label{fig:three_edges}
\end{figure*}

We first characterized the broadband HETG continuum with a two-component thermal model consisting of a soft disk blackbody (\texttt{dbb}) and a harder blackbody (\texttt{bb}). This approach follows previous studies of \gx\ and other \emph{Z} sources \citep[e.g.,][]{dAi09,Bhargava23}. Although broadband spectra of \gx\ often include an additional non-thermal component, the restricted HETG bandpass does not allow its contribution to be constrained. \astrosat\ observations indicate that the non-thermal emission becomes significant mainly above 10~keV \citep[below $\sim1.24$~\AA;][]{Bhargava23}. The intrinsic emission was attenuated by neutral interstellar absorption using the \texttt{hot} model \citep{dePlaa04}. Elemental abundances were adopted from \citet{Lodders09}; the column density was initialized at $N_{\rm H}=10^{23}\ \rm cm^{-2}$ and left free to vary. The \texttt{hot} temperature was fixed at its lower limit ($kT=10^{-6}$~keV). The distance was fixed to 11~kpc \citep{Penninx93}. The cross-normalization between MEG and HEG was allowed to vary and remained within 3\%. The baseline model \texttt{hot*(dbb+bb)} yields $\cstat/\rm dof = 2789/1642$, where dof denotes the number of degrees of freedom (Figure~\ref{fig:continuum_residuals}). Tests with alternative continua confirmed that the inferred local edge structures, which are the primary focus of this work, are not sensitive to the exact continuum prescription.

To describe the Fe~K emission, we included both a broad relativistic component and a narrower line. The broad profile was constructed by convolving a delta function (\texttt{delt}) with the relativistic line kernel \texttt{spei} \citep{Speith95}, and an additional narrow Gaussian (\texttt{gaus}) was used to capture unresolved structure. Initial values for the inner radius, inclination, emissivity index, and line energy were guided by \citet{dAi09} and \citet{Ludlam25}. This phenomenological reflection prescription is intentionally simpler than the self-consistent relativistic reflection modeling adopted in \citet{Ludlam25}, but it is sufficient for our purposes of describing the Fe~K emission while focusing on the absorption edges. Following previous work, the Fe abundance in the absorbing ISM component was allowed to vary to account for the deep Fe~K edge at 7.1~keV. This model yields $\cstat/\rm dof = 2457/1633$.

Significant residuals near the Si~K edge around 6.74~\AA\ (Figure~\ref{fig:continuum_residuals}) indicate the presence of interstellar dust along the line of sight \citep{Zeegers17,Zeegers19,Rogantini20}. To model the X-ray absorption fine structure (XAFS) imprinted by grains, we included the \spex\ dust extinction component \texttt{amol} \citep{Pinto10}. When dust is added, the gas-phase abundances of elements that contribute significantly to solids (O, Mg, Si, S, and Fe) must be adjusted to account for depletion into the solid phase; otherwise the fit can drive the total abundances to unphysical super-solar values. We therefore parameterize each element with a gas-phase fraction, $\delta_X \equiv A_{X,\rm gas}/A_{X,\rm total}$ (so that the dust-phase
fraction is $1-\delta_X$), and restrict these fractions to physically motivated ranges based on depletion studies and previous X-ray work \citep{Whittet02,Jenkins09,Rogantini20,Psaradaki23}: $\delta_{\rm Fe}=0$--0.2, $\delta_{\rm Mg}=0$--0.6, $\delta_{\rm Si}=0$--0.6, and $\delta_{\rm O}=0.4$--0.8. We verified that widening these bounds does not change the inferred dust parameters within uncertainties. For sulfur, the \texttt{hot} abundance is fixed to zero once the \Sii\ opacity is modeled explicitly with \texttt{musr} (see below). The fitted oxygen and magnesium fractions reach their imposed upper limits (0.8 and 0.6, respectively). Because the O and Mg K edges lie outside the HETG wavelength range considered here, we cannot directly constrain their depletion (and thus their dust-to-gas ratios); the high column density also prevents measuring these edges with other high-resolution instruments, such as the Reflection Grating Spectrometer onboard \xmm. We therefore do not attempt to interpret the inferred O and Mg depletion further in the discussion.

A narrow Gaussian line was also included to model the known instrumental feature\footnote{\url{https://space.mit.edu/CXC/calib/sikedge_final_doc.pdf}} at 6.741~\AA, which overlaps the \sii\ line region near the Si~K edge \citep{Rogantini20,Yang22}. A preliminary fit including three representative dust species (a silicate, metallic iron, and an iron sulfide) yields $\cstat/\rm dof = 2236/1628$. While the dust extinction model reproduces the Si~K-edge fine structure well, residuals remain at the S~K edge. Two prominent absorption-line residuals appear in the HEG spectrum, consistent with low-ionization sulfur. To fit these relevant low-ionization sulfur transitions, we incorporated sulfur photoabsorption cross sections from \citet{Gatuzz24a} via the user-defined multiplicative model \texttt{musr} in our local \spex\ installation, following the approach adopted by \citet{Corrales25}. Specifically, the S~\textsc{ii} cross section was added and its column density was fitted freely. To avoid driving the total sulfur abundance to unphysical values when including this additional sulfur opacity, the sulfur abundance in the \texttt{hot} component was fixed to zero. This significantly improves the S~K-edge fit, resulting in a final $\cstat/\rm dof = 2006/1626$. The best fit requires an additional velocity shift of the sulfur cross section of $977\pm10$~km~s$^{-1}$ (corresponding to an energy shift of 8.05~eV). A comparable shift of $\sim$8~eV was also required by \citet{Corrales25} when fitting the S~K edge in the \xrism\ spectrum.

The stacked HETG spectrum provides high signal-to-noise around the Si~K edge. A systematic shift is observed between the Si~K-edge position in the data and the laboratory-based Si-bearing dust models. To quantify this, the Si~K edge was fitted independently while allowing the \texttt{amol} velocity parameter ($zv$) to vary. The inferred shift is $-206\pm43$~km~s$^{-1}$, corresponding to an energy offset of $\sim$1.3~eV, comparable to the typical laboratory calibration uncertainty \citep{Zeegers19}. A similar offset was previously reported for GX~3+1 \citep{Rogantini19}. In the deep stacked spectrum of \gx, the large optical depth of the Si~K edge allows this shift to be measured at $>20\sigma$ significance. We therefore use the HEG spectrum to recalibrate the Si-bearing dust models adopted in this work.

Possible additional components were also explored, motivated by the \xrism\ results, such as ionized absorption from an accretion-disk wind and photoionized emission from ionized gas \citep[e.g.,][]{Chakraborty25}. We added a photoionized absorber and a photoionized emission component (both modeled with \texttt{pion}; \citealt{Mehdipour16}), but neither improved the fit significantly. At the \chandra/HETG spectral resolution in the Fe~K band, these features are likely blended with the relativistic emission profile and cannot be robustly separated. The final adopted model is therefore \texttt{hot*amol*musr*(dbb+bb+gaus+delt*spei) + gaus}, where the additive Gaussian represents the instrumental feature at 6.741~\AA.

\begin{figure*}[t]
\centering
\includegraphics[width=\textwidth]{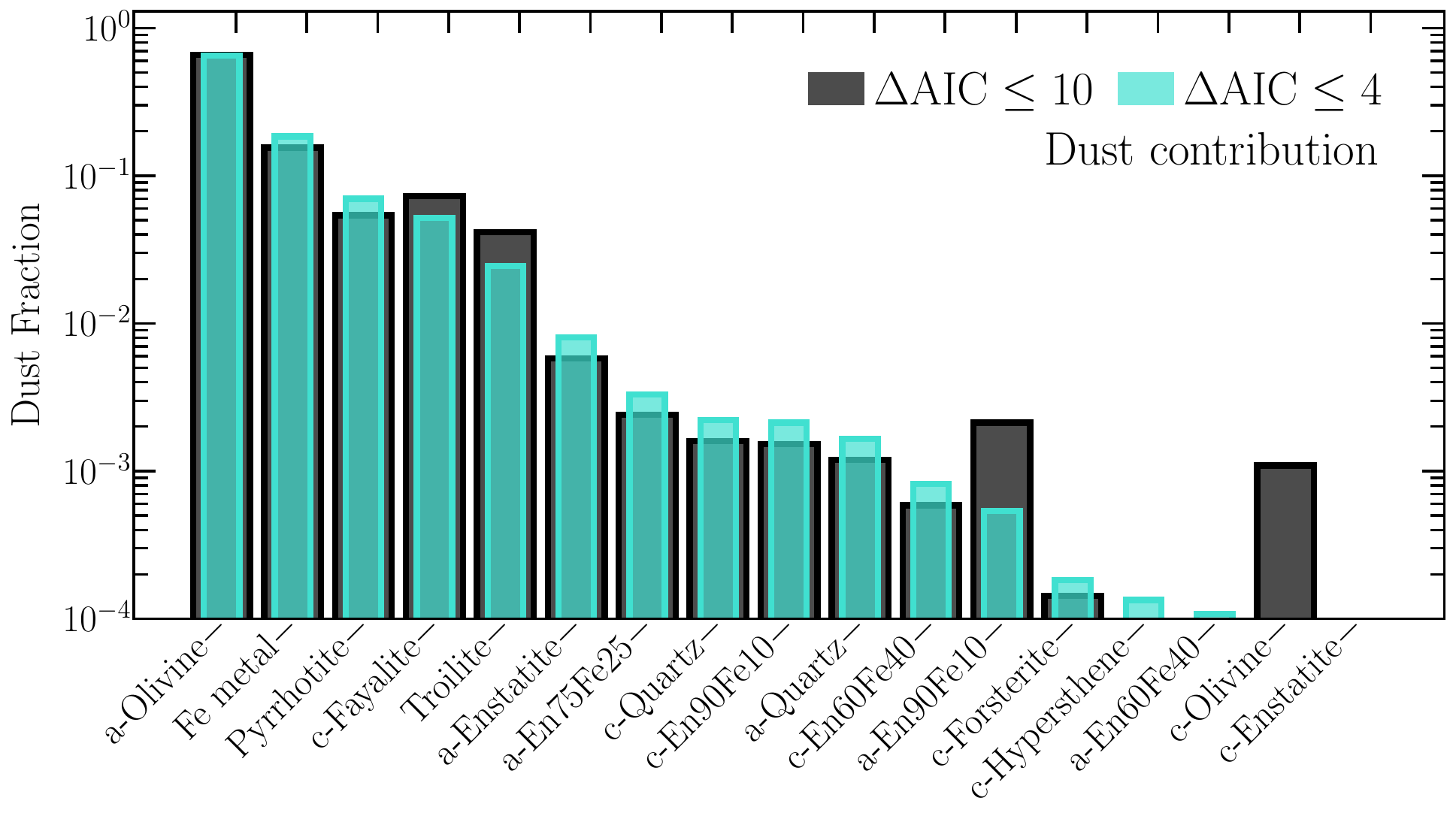}\\
\caption{Relative fraction of the dust species obtained from models with $\Delta\mathrm{AIC}\leq4$ (light blue) and $\Delta\mathrm{AIC}\leq10$ (dark grey). The prefixes c- and a- denote crystalline and amorphous forms, respectively. A detailed description of the dust extinction models is provided in \citet{Zeegers19,Rogantini19,Costantini19}. Further information is available on the \spex\ website page dedicated to the \texttt{amol} model.}
\label{fig:dust_fraction}
\end{figure*}

The \texttt{amol} database contains several dozen dust compounds. From this set, 17 species relevant to the Si, S, and Fe K edges were selected (Figure~\ref{fig:dust_fraction}). To explore the dust composition while keeping the number of free parameters manageable, combinations of up to four dust species were fitted at a time, corresponding to the maximum number allowed within a single \texttt{amol} component. For the 17 candidate compounds shown in Figure~\ref{fig:dust_fraction}, this yields ${17 \choose 4}=2380$ unique dust mixtures. For each mixture, the full model was refitted and the Akaike Information Criterion was computed, $\mathrm{AIC}=2k+\cstat$, where $k$ is the number of free parameters. The best-fit parameters and their $1\sigma$ uncertainties for the preferred dust model, that is, the combination with the lowest C-statistic and lowest AIC, are listed in Table~\ref{tab:results}. Figure~\ref{fig:three_edges} shows the corresponding best fit in the Si, S, and Fe K-edge regions. To assess which of the tested models carry comparable statistical support, two statistically selected ensembles were defined: a conservative set with $\Delta\mathrm{AIC}\leq4$ (36 models) and a broader set with $\Delta\mathrm{AIC}\leq10$ (52 models), following \citet{Burnham02} and previous X-ray dust studies. Models with $\Delta\mathrm{AIC}\leq4$ are considered statistically indistinguishable from the best-fit model, while models with $\Delta\mathrm{AIC}\leq10$ cannot be confidently excluded \citep[for a detailed methodology, see][]{Rogantini19}. The average dust properties of each ensemble were then investigated.

\begin{table}[t]
\caption{Spectral analysis results for the stacked \chandra/HETG spectra of \gx: best-fit values for the continuum, emission, and absorption components.}
\label{tab:results}
\centering
\hspace*{-\dimexpr\oddsidemargin+0.4in}
\begin{threeparttable}
\begin{tabular}{c c | c l }
\hline\hline
Comp. & Par. & Value & Units \\
\hline
\noalign{\vskip 0.1mm}
\multicolumn{4}{c}{Neutral interstellar absorption} \\
\noalign{\vskip 0.1mm}
\hline
\multirow{7}{*}{\texttt{hot}} & $N_{\rm H}$ & $6.89_{-0.11}^{+0.10}$ & $10^{22}\ \rm cm^{-2}$ \\
  & $kT$  & $10^{-6}$ & keV \\
  & $\delta_{\rm O}^{\dagger}$  & $0.80^{f}$ & \\
  & $\delta_{\rm Mg}^{\dagger}$ & $0.60^{f}$ & \\
  & $\delta_{\rm Si}^{\dagger}$ & $0.041\pm0.009$ & \\
  & $\delta_{\rm S}^{\dagger}$  & $0.00^{f}$ & \\
  & $\delta_{\rm Fe}^{\dagger}$ & $0.0^{+0.2}_{-0.0}$ & \\
\hline
\noalign{\vskip 0.1mm}
\multicolumn{4}{c}{Continuum} \\
\noalign{\vskip 0.1mm}
\hline
\multirow{2}{*}{\texttt{dbb}} & Norm & $2.88_{-0.07}^{+0.11}$ & $10^{11}\ \rm cm^{2}$ \\
 & $kT^{\ddagger}$ & $3.85_{-0.04}^{+0.02}$ & keV \\
\hline
\multirow{2}{*}{\texttt{bb}} & Norm & $8.4_{-0.4}^{+0.11}$ & $10^{10}\ \rm cm^{2}$ \\
 & $kT$ & $5.80_{-0.07}^{+0.06}$ & keV \\
\hline
\noalign{\vskip 0.1mm}
\multicolumn{4}{c}{Fe K emission} \\
\noalign{\vskip 0.1mm}
\hline
\multirow{3}{*}{\texttt{gaus}} & Norm & $0.11_{-0.02}^{+0.01}$ & $10^{44}\ \rm ph\,s^{-1}$ \\
& $\lambda$ & $1.856\pm0.003$ & \AA \\
& $\rm FWHM$ & $0.04\pm0.01$ & \AA \\
\hline
\multirow{2}{*}{\texttt{delt}} & Norm & $1.21_{-0.10}^{+0.11}$ & $10^{44}\ \rm ph\,s^{-1}$ \\
& $\lambda$ & $1.790^{+0.012}_{-0.006}$ & \AA \\
\hline
\multirow{4}{*}{\texttt{spei}} & $r_{\rm in}$ & $17_{-1}^{+10}$ & $\rm GM/c^{2}$ \\
& $r_{\rm out}$ & $460_{-90}^{+150}$ & $\rm GM/c^{2}$ \\
& $i$ & $55\pm5$ & degree \\
& $q$ & $2.4\pm0.1$ & \\
\hline
\noalign{\vskip 0.1mm}
\multicolumn{4}{c}{Interstellar dust} \\
\noalign{\vskip 0.1mm}
\hline
\multirow{4}{*}{\texttt{amol}} & $N_{1}$ & $10^{+2}_{-5}$ & $10^{17}\ \rm cm^{-2}$ (Metallic Fe) \\
& $N_{2}$ & $3.4_{-1.0}^{+0.9}$ & $10^{17}\ \rm cm^{-2}$ (Pyrrhotite) \\
& $N_{3}$ & $<1$ & $10^{17}\ \rm cm^{-2}$ (a-Quartz) \\
& $N_{4}$ & $24\pm1$ & $10^{17}\ \rm cm^{-2}$ (a-Olivine) \\
\hline
\noalign{\vskip 0mm}
\multicolumn{4}{c}{Instrumental line} \\
\noalign{\vskip 0mm}
\hline
\multirow{3}{*}{\texttt{gaus}} & Norm & $0.006_{-0.001}^{+0.002}$ & $10^{44}\ \rm ph\,s^{-1}$ \\
& $\lambda$ & $6.740\pm0.001$ & \AA \\
& $\rm FWHM$ & $<0.002$ & \AA \\
\hline
\noalign{\vskip 0mm}
\multicolumn{4}{c}{Sulfur photoelectric absorption} \\
\noalign{\vskip 0mm}
\hline
\texttt{musr} & $N_{\rm S\,\textsc{ii}}$ & $6.4_{-0.9}^{+1.0}$ & $10^{17}\ \rm cm^{-2}$ \\
\hline
\noalign{\vskip 0mm}
\multicolumn{4}{c}{Flux estimates} \\
\noalign{\vskip 0mm}
\hline
\multicolumn{2}{c|}{$F_{\rm observed;\ 2\mbox{-}10\ \rm keV}$}  & $8.05\pm0.75$ & $10^{-9}\ \rm erg\,s^{-1}\,cm^{-2}$ \\
\multicolumn{2}{c|}{$L_{2\mbox{-}10\ \rm keV}$}  & $1.90\pm0.20$ & $10^{38}\ \rm erg\,s^{-1}$ \\
\hline
\multicolumn{2}{c|}{$C$stat/dof} & 2006/1626 \\
\hline
\end{tabular}
\begin{tablenotes}
\item $^{f}$ Frozen parameters.
\item $^{\dagger}$ Gas-phase abundance fractions in \texttt{hot}.
\item $^{\ddagger}$ The definition of the disk-blackbody temperature differs between \spex\ and \xspec; \spex\ \texttt{dbb} temperatures are approximately twice as high.
\end{tablenotes}
\end{threeparttable}
\end{table}

%The best-fit parameters and their $1\sigma$ uncertainties for the preferred dust model are listed in Table~\ref{tab:results}. Figure~\ref{fig:three_edges} shows the corresponding best fit in the Si, S, and Fe K-edge regions. Given the large number of dust combinations explored, model comparison was performed using the Akaike Information Criterion \citep[AIC;][]{Akaike74}. For Cash statistics, the AIC can be written as $\mathrm{AIC}=2k+\cstat$, where $k$ is the number of free parameters. Following \citet{Burnham02}, 

Figure~\ref{fig:dust_fraction} summarizes the dust composition inferred from the statistically selected ensembles. For each model in a given ensemble, the fitted dust column densities are converted into fractional contributions by normalizing to the total dust column density of that model, i.e., $f_j = N_j/\sum_i N_i$ for dust species $j$. The ensemble-averaged fraction of each compound is then obtained by assigning zero contribution to species not included in a given model and averaging the resulting fractions across the full ensemble. Unless otherwise stated, the quoted dust fractions and derived Fe partition are based on the $\Delta\mathrm{AIC}\le4$ ensemble; the corresponding values for $\Delta\mathrm{AIC}\le10$ are similar and are shown for comparison in Figure~\ref{fig:dust_fraction}.

Two complementary sets of quantities are reported in this work: (1) parameters derived from the single best-fit dust model, and (2) quantities derived from the ensemble of statistically similar models. The elemental gas and dust columns, dust fractions per element, and abundances in Table~\ref{tab:abundances} are computed from the best-fit model, i.e., the minimum-AIC solution. Note that the $A/A_\odot$ values are derived from the total elemental columns, $N_{\rm gas}+N_{\rm dust}$, and therefore do not correspond directly to the fitted \texttt{hot} abundance parameters listed in Table~\ref{tab:results}. In particular, for sulfur, $N_{\rm gas}$ is provided by the \texttt{musr} component (S~\textsc{ii}), while the sulfur abundance in \texttt{hot} is fixed to zero to avoid double-counting the sulfur opacity.

By contrast, the dust-species fractions shown in Figure~\ref{fig:dust_fraction}, as well as the inferred partition of iron among silicates, sulfides, and metallic iron, are derived from the ensemble of statistically acceptable models. This approach captures the uncertainty introduced by degeneracies among dust mixtures, especially at the S and Fe K edges, where multiple combinations can provide similarly good fits while slightly changing the inferred distribution among individual compounds.

In the restricted set of statistically acceptable models ($\Delta\mathrm{AIC}\le4$), amorphous olivine ($\rm MgFeSiO_4$) dominates the inferred dust mixture, contributing $\sim65\%$ of the total dust column density and appearing in all selected combinations. Metallic iron contributes $\sim19\%$ on average and is included in $\sim64\%$ of the selected models, reflecting degeneracy at the Fe~K edge. Iron sulfides account for $\sim10\%$ of the dust budget: pyrrhotite ($\rm Fe_{0.875}S$) is present in $\sim70\%$ of the selected models, and when pyrrhotite is not included, troilite (FeS) typically provides the preferred sulfide contribution, effectively replacing it. Fayalite ($\rm Fe_{2}SiO_4$) contributes $\sim5\%$ in this restricted set. The remaining species contribute only a few percent in total, with each individual compound contributing $<1\%$.

To estimate how Fe is distributed among silicates, sulfides, and metallic form, we weight each dust-species fraction by the number of Fe atoms in its stoichiometric formula and renormalize to the total Fe locked in the selected dust species. We notice that gas-phase iron fraction is degenerate with the metallic iron dust column density. In models including metallic iron, the gas-phase iron fraction converges to zero. Conversely, in models without metallic iron, the gas-phase fraction varies between 0.1 and 0.2, often reaching the imposed upper limit. This degeneracy is driven by the limited HETG energy resolution in the Fe~K band, which does not allow the absorption cross sections of gaseous and metallic iron to be cleanly separated. Consequently, both scenarios provide statistically comparable fits to the Fe~K edge.

\section{Discussion} \label{sec:discussion}

\begin{deluxetable}{lcccc}
\tablecaption{Gas and dust columns for Si, S, and Fe along the line of sight to \gx\ for the best-fit model. Columns are reported in units of $10^{17}\,\rm cm^{-2}$. The dust fraction is defined as $f \equiv N_{\rm dust}/(N_{\rm gas}+N_{\rm dust})$. Abundances are computed as $A/A_{\odot}=(N_{\rm gas}+N_{\rm dust})/N_{\odot}$ using the reference solar columns from \citep{Lodders09}. \label{tab:abundances}}
\tablehead{
\colhead{Element} & \colhead{$N_{\rm gas}$} & \colhead{$N_{\rm dust}$} & \colhead{$f$} & \colhead{$A/A_{\odot}$}
}
\startdata
Si & $1.1\pm0.2$ & $24_{-2}^{+1}$ & $0.96\pm0.01$ & $0.94\pm0.05$ \\
S  & $6.4^{+1.0}_{-0.9}$ & $3.4^{+0.9}_{-1.1}$ & $0.35^{+0.07}_{-0.08}$ & $0.87\pm0.12$ \\
Fe & $<4.50$ & $37^{+3}_{-6}$ & $>0.89^{a}$ & $1.66^{+0.12}_{-0.21}\ (<1.9)^{b}$ \\
\enddata
\tablecomments{
$^{a}$ For Fe, $N_{\rm gas}$ is constrained only by an upper limit; therefore $f_{\rm Fe}$ is reported as a lower limit. The best-fit solution has $N_{\rm gas,Fe}\simeq 0.03 \times10^{17}\ \mathrm{cm^{-2}}$, implying $f_{\rm Fe}\approx 1$.
$^{b}$ The value outside parentheses uses $N_{\rm dust}$ only since $N_{\rm gas}=0$ for the best fit; the upper limit in parentheses includes the maximum allowed gas contribution.
Dust columns exclude the a-quartz (SiO$_2$) component, which is consistent with zero within uncertainties; its contribution is negligible and does not affect the derived quantities within the quoted uncertainties.}
\end{deluxetable}

\begin{figure}[t]
\centering
\includegraphics[width=\columnwidth]{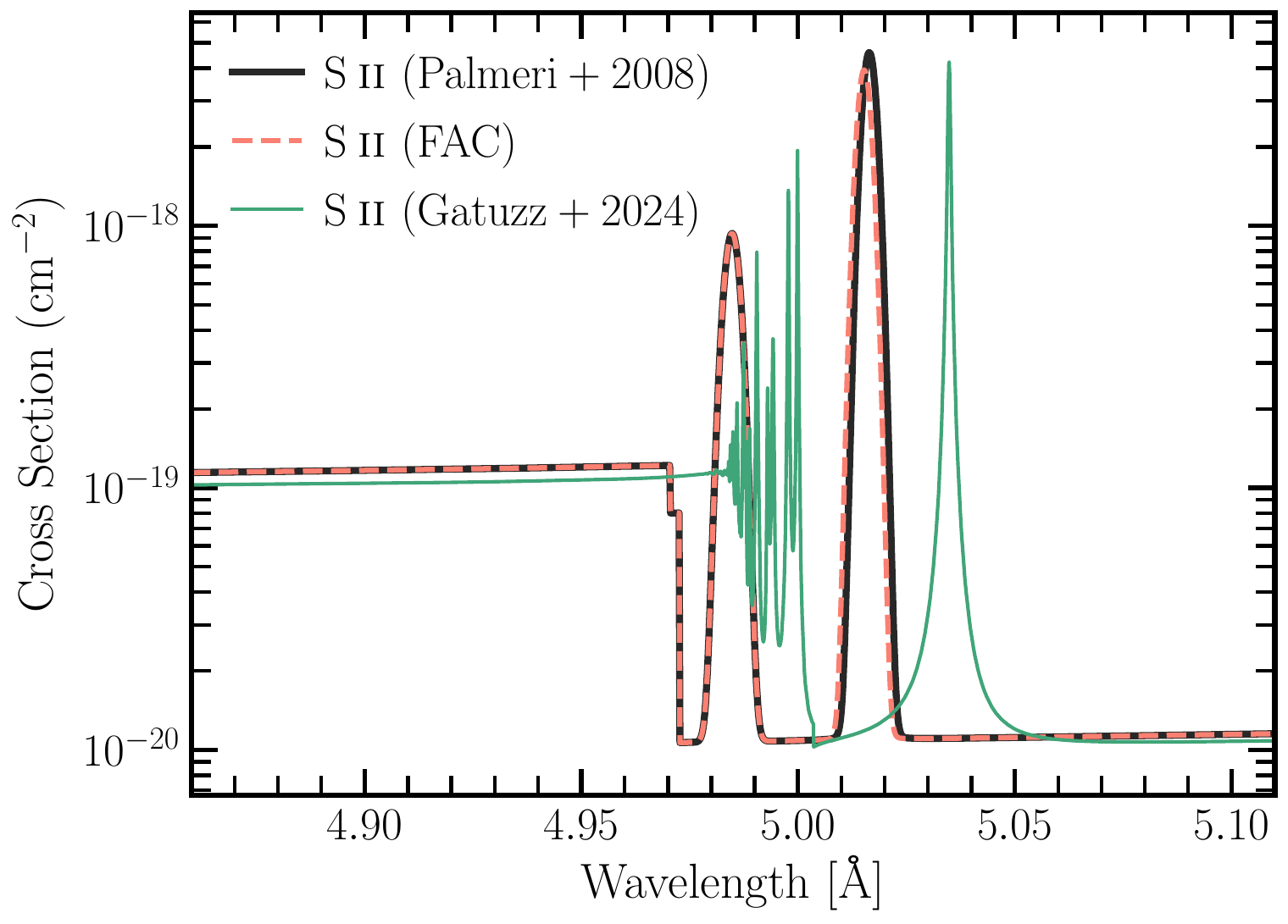}\\
\caption{Comparison of S~\textsc{ii} photoabsorption cross sections adopted to model the S~K edge. Shown are the cross section from \citet{Palmeri08} augmented with the K$\gamma$ line from the FAC calculation, the full FAC-based cross section implemented in \spex, and the $R$-matrix cross section from \citet{Gatuzz24a}}. %The positions of the K$\beta$ and K$\gamma$ transitions are marked.}
\label{fig:sii}
\end{figure}

\begin{figure*}[t]
\centering
\includegraphics[width=\textwidth]{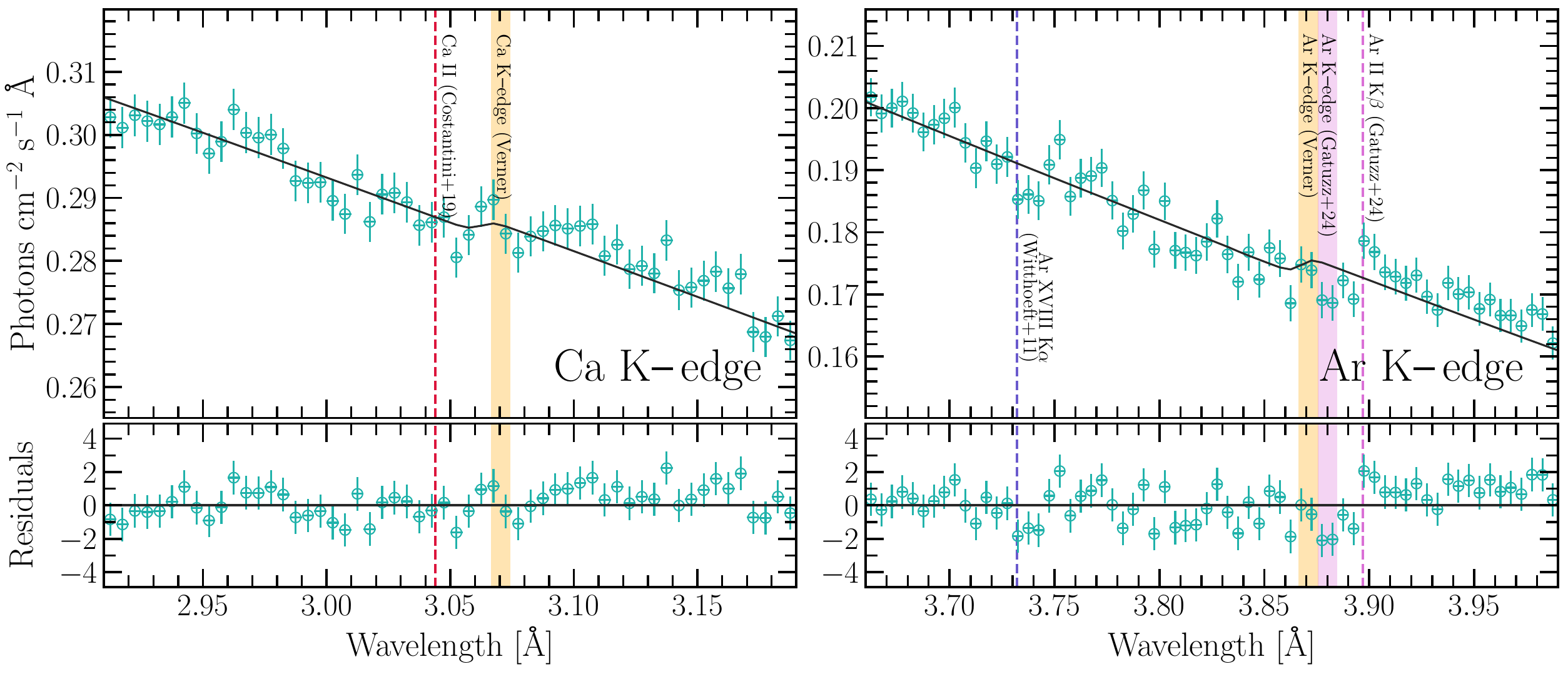}\\
\caption{Detection of low-abundance metals in the stacked HEG spectrum of \gx, showing the Ca and Ar K-edge regions. Key spectral features from \citet{Verner96}, \citet{Costantini19}, \citet{Witthoeft11}, and \citet{Gatuzz24b} are marked. The residuals are computed with respect to the best-fit model shown in Figure \ref{fig:three_edges} and described in Table \ref{tab:results}.}
\label{fig:calcium_argon}
\end{figure*}

We simultaneously modeled the Si, S, and Fe K edges in the HETG spectra of \gx to constrain the properties of the interstellar medium and dust along this highly absorbed line of sight. Below, we discuss the implications of the best-fit model and the resulting dust composition.

\subsection{Silicon}
The stacked HEG spectrum provides high-quality data at the Si~K edge, where the HETG resolving power peaks (energy resolution $\simeq0.01$~\AA, corresponding to $\sim$2.5~eV at 1.84~keV). This enables the direct detection of X-ray absorption fine structure (XAFS) and therefore tight constraints on the composition and crystallinity of Si-bearing grains. At the Fe and S K edges the spectral resolution is lower, limiting sensitivity to fine structure. A simultaneous fit of the Si, S, and Fe K edges is therefore essential to determine the full dust composition and to break degeneracies in the dust modeling: the Si~K-edge XAFS constrains silicate chemistry and crystallinity, while the optical depths at the S and Fe K edges refine the contribution from non-silicate compounds.

Our fits indicate that amorphous olivine dominates the dust composition, contributing $65\%$ of the total dust column density, and it is present in all statistically selected model combinations. This is consistent with \citet{Zeegers19}, who analyzed the single Si~K edge in a subset of HETG observations of \gx (150~ks total) and found that amorphous olivine contributes 70--90\% of the Si~K-edge absorption. \citet{Yang22} also studied the Si~K edge of \gx\ and found that olivine provides the best fit; however, their dust sample did not include an amorphous olivine analog, only crystalline counterparts. Table~\ref{tab:abundances} shows a near-solar silicon abundance ($A_{\rm Si}/A_\odot\simeq1$) and a high dust fraction, with $f_{\rm Si}\simeq0.96$ (i.e., $\sim$96\% of Si locked into solids). Overall, our results provide strong evidence that $\rm SiO_4$ olivine-type silicates are preferred over $\rm SiO_3$ pyroxene-type silicates along this highly-absorbed sightline, consistent with findings for other high-column-density lines of sight \citep{Zeegers19,Rogantini20}. Among pyroxenes, amorphous enstatite is the most prominent, but its contribution remains below $\sim2\%$ of the total dust column. In contrast, more diffuse sightlines tend to favor Mg-rich pyroxenes, suggesting a different dust population \citep{Psaradaki23}.

\subsection{Iron}
Within the $\Delta\mathrm{AIC}\le4$ set, Fe-bearing silicates (amorphous olivine, $\rm MgFeSiO_4$, and fayalite, $\rm Fe_2SiO_4$) account for $\sim74\%$ of the dust-phase Fe, while Fe sulfides contribute $\sim8\%$. The remaining $\sim18\%$ is associated with metallic iron. Given the limited HETG resolving power at the Fe~K edge and the partial degeneracy between metallic and gas-phase Fe, these fractions should be regarded as model-dependent estimates rather than unique identifications. The dust-model preference is driven mainly by the Si~K edge, which has the highest signal-to-noise ratio and the strongest XAFS signatures. Our preference for Fe-rich silicates is qualitatively consistent with mid-IR studies of stellar outflows. For example, \citet{Tamanai17} found that reproducing silicate band profiles in supergiant spectra requires amorphous silicates with substantial iron content (Mg:Fe $\sim$1), and that metallic iron particles may also be present in outflows; their models disfavour iron-poor silicates.

At lower column densities ($\NH\lesssim10^{21}\ \rm cm^{-2}$), Fe~L-edge spectroscopy has been used to constrain the iron budget. \citet{Westphal14} analyzed Cyg~X-1 and found that $\lesssim39\%$ of Fe is in silicates, with most iron in metallic form and no significant FeS contribution. \citet{Corrales24} modeled the Fe~L edges of Cyg~X-1 and GX~339--4 and found that iron oxides and fayalite provide comparably good fits, although a fayalite-dominated model was disfavored based on Mg constraints. \citet{Psaradaki23} studied Fe~L and O~K edges in five LMXBs and concluded that $\sim40\%$ of Fe is metallic and $\sim60\%$ is in silicates, with $<2\%$ in iron sulfides.

Because the HETG energy resolution is limited in the Fe~K band, metallic and gaseous iron are difficult to distinguish spectroscopically, as both primarily contribute only through the Fe~K edge. Nevertheless, Table~\ref{tab:abundances} indicates a high dust fraction for iron ($f_{\rm Fe}>0.87$) and a super-solar total iron abundance ($A_{\rm Fe}/A_\odot\simeq1.7$, with an upper limit of $<2.0$ when including the maximum allowed gas contribution). This is consistent with depletion studies indicating that in denser regions a very large fraction of iron ($\gtrsim95\%$) should reside in dust \citep[e.g.,][]{Zhukovska18}.

Our results indicate a relatively smaller metallic-Fe fraction than some Fe~L-edge studies, and are broadly consistent with constraints from sub-mm polarization. If metallic Fe exists as inclusions within grains, it can contribute to grain alignment and to thermal magnetic dipole emission \citep{Draine99,Draine13}. \citet{Draine13} argue that the fraction of Fe in ferromagnetic form cannot be too large if the phenomenological Gilbert equation applies, because the resulting frequency dependence of polarized emission would conflict with \emph{Planck} constraints and would produce polarized anomalous microwave emission (AME) from molecular clouds, whereas AME is typically associated with the diffuse ISM \citep[e.g.,][]{Dickinson18,Hensley23}. A larger fraction of Fe could be in ferromagnetic form if the Gilbert equation does not accurately describe the magnetic behaviour at very high frequencies (Bruce Draine, priv.\ comm.; \citealt{Hensley23}).

A key open question remains the chemical form of the remaining Fe outside silicates. Current and future calorimeter spectroscopy will tighten constraints in the Fe~K band. The existing \xrism/\textsc{Resolve} observation of \gx already demonstrates the potential of this approach, and deeper calorimeter data would further constrain both the Fe~K-edge structure and the complex Fe~K emission \citep[e.g.,][Psaradaki et al.\ in prep.]{Ludlam25,Chakraborty25}. At the same time, \chandra/HETG\ remains crucial for dust chemistry thanks to its high and relatively uniform resolving power at the Mg, Si, and S K edges, which are essential for separating silicate and sulfide contributions. Complementary \chandra/LETG\ and HETG/HRC observations can also access the O~K and Fe~L edges at lower energies. In parallel, we are expanding the available iron dust models by adding new iron extinction cross sections to the \textsc{amol} database (Zeegers et al.\ in prep.).

\subsection{Sulfur}
The high column density and long exposure enable a significant detection of the S~K edge. Dust extinction alone cannot reproduce the detailed structure: two strong absorption features overlap the edge region near 4.98~\AA\ and 5.02~\AA, consistent with the \Sii\ K$\beta$ and K$\gamma$ transitions. To model these features, we adopted the \Sii\ photoabsorption cross section computed with the $R$-matrix method \citep{Berrington95,Burke11} and implemented it via the \texttt{musr} model in \spex\ \citep{Gatuzz24a}. This treatment includes Auger broadening and provides a substantially improved description of the edge region.

Our best-fit models favor a significant fraction of sulfur in Fe-bearing sulfides, with pyrrhotite emerging as the preferred carrier among the tested species. We infer a sulfur dust fraction of $f_{\rm S}\simeq0.35$ (i.e., $\sim35\%$ of sulfur in the condensed phase) along this highly-absorbed sightline. The total sulfur abundance is sub-solar but consistent with unity within uncertainties, with $A_{\rm S}/A_\odot = 0.87\pm0.12$ (Table~\ref{tab:abundances}). This result is consistent with the general picture in which sulfur is largely undepleted in diffuse environments but becomes increasingly incorporated into the condensed phase in denser regions \citep{Keller02,Scappini03,Laas19,Jenkins09,Ferrari24}. While sulfur may also reside in organic compounds on grains \citep{Laas19}, such species are not included in current X-ray dust extinction databases and cannot be directly tested with our modeling. However, Fe sulfides are commonly observed in primitive solar-system materials and in GEMS \citep{Bradley94}.

Our sulfur dust fraction is consistent with the \xrism/\textsc{Resolve} analysis of \gx\ by \citet{Corrales25}, who also find that $\sim40\%$ of sulfur is in the condensed phase when modeling the S~K edge with comparable gas and dust cross sections. \citet{Gatuzz24a} analyzed the S~K edge in a larger sample of X-ray binaries (including \gx) and reported \Sii\ and \Siii\ column densities of order $\sim2\times10^{17}\ \rm cm^{-2}$, while placing stringent upper limits on sulfur-bearing dust columns in lower-column sightlines. Joint far-UV and X-ray studies also support \Sii\ as the dominant gas-phase ion: \citet{Psaradaki24} measured \Sii\ absorption in Cyg~X-2 at 1250.6~\AA\ and 1253.8~\AA\ and, using \texttt{Cloudy} \citep{Ferland17}, concluded that \Sii\ dominates over \Si\ and \Siii, accounting for $\sim35\%$ of the total sulfur for solar abundances, with the remainder likely bound in molecules or dust.

The inferred sulfur gas-to-dust partition depends sensitively on the adopted \Sii\ atomic data. Figure~\ref{fig:sii} compares several \Sii\ photoabsorption cross sections available in the literature and those implemented in \spex. The $R$-matrix calculation of \citet{Gatuzz24a} provides a close-coupling treatment of photoionization that includes channel coupling and orbital relaxation, yielding accurate resonance positions and edge structure; Auger broadening is treated self-consistently. In contrast, the cross section of \citet{Palmeri08} was computed with a pseudo-relativistic Hartree--Fock method, which offers an efficient description of inner-shell photoabsorption but can predict different resonance energies and edge positions, particularly when relaxation effects are important. Calculations based on the Flexible Atomic Code (FAC), a widely used atomic-structure and collision package, were also considered; FAC computes level energies and transition rates using a relativistic (Dirac) mean-field approach. FAC-based cross sections are implemented in \spex\ and can be used to supplement missing transitions (e.g., the K$\gamma$ line absent in \citealt{Palmeri08}). For \Sii, FAC wavelengths are typically accurate to $\sim$0.01--0.02~\AA, while absorption line strengths carry $\sim$20\% uncertainties (Ming-Feng Gu, priv.\ comm.). These different approaches lead to measurable shifts in the energies of the main resonances and in the edge position, as also highlighted by \citet{Gatuzz24a} when comparing their $R$-matrix results to earlier datasets used in spectral modeling \citep[e.g.,][]{Witthoeft09,Witthoeft11}. While the overall shapes of the \Sii\ cross sections are broadly similar, the inclusion of orbital relaxation in the $R$-matrix calculation shifts both the edge position and the resonance energies.

We tested the \citet{Palmeri08}-based profile (augmented with the K$\gamma$ line from FAC), the full FAC-based prescription used in \spex, and the $R$-matrix cross section of \citet{Gatuzz24a}. We adopted the $R$-matrix results because they are strongly favored statistically ($\Delta$Cstat$=25$ relative to the alternative implementations, in particular the full FAC-based one). Using different \Sii\ cross sections does not significantly change the inferred dust composition or dust column densities, but it does affect the derived \Sii\ column density and therefore the sulfur dust-to-gas ratio. For example, when adopting the full FAC cross section (which is similar in shape to the \citealt{Palmeri08} profile), we obtain a substantially smaller gaseous column density and comparable dust contribution leading to a dust fraction of $N_{\rm dust}/(N_{\rm gas}+N_{\rm dust})\simeq0.79$ (i.e., $\sim79\%$ of sulfur in dust) illustrating the size of the systematic uncertainty associated with the adopted atomic data. While systematic effects driven by atomic-data uncertainties cannot be fully excluded, we adopt the \citet{Gatuzz24a} cross section as our reference because it provides the best statistical fit ($\Delta C_{\rm stat}=25$ for the same number of degrees of freedom) and yields a sulfur depletion more consistent with expectations from multiwavelength studies \citep[e.g.,][]{Jenkins09}.

\subsection{Low abundance elements: argon and calcium}

The exceptional quality of the stacked HEG spectrum reveals absorption structure in the Ar~K-edge region (right panel of Figure~\ref{fig:calcium_argon}). As a noble gas, argon is expected to remain in the gas phase in the ISM. In our spectrum, the commonly used Verner cross section (yellow band) does not reproduce the observed edge shape, while an edge-like feature coincides with the Ar~\textsc{ii} K$\beta$ resonance computed by \citet{Gatuzz24b} using the $R$-matrix method. \citet{Gatuzz24b} modeled the Ar~K edge in a large sample of X-ray binaries, including photoabsorption cross sections for Ar~\textsc{i--iii} and Ar~\textsc{xvi--xviii}, and reported only upper limits for \gx\ (except for Ar~\textsc{xviii}). We searched for a hot ISM contribution by adding an additional \texttt{hot} component with $kT>0.05$~keV, but found no statistically significant improvement. Overall, the Ar~K-edge region in \gx\ highlights the need for improved atomic photoabsorption data; the observed edge structure appears offset with respect to some commonly adopted cross sections. This will be particularly important for current (\xrism) and future (e.g. {\it NewAthena}) microcalorimeter observations, whose higher spectral resolution at 3--4~keV will enable a more precise characterization of the Ar~K-edge structure.

The Ca~K edge is only marginally detected (left panel of Figure~\ref{fig:calcium_argon}). Calcium is expected to be strongly depleted in the ISM \citep{Crinklaw94}, primarily residing in silicates and oxides \citep{Field74,Trivedi81,ReyMontejo24}. The combination of its small optical depth and the limited \chandra/HETG\ resolving power at short wavelengths prevents a detailed study of the Ca~K-edge fine structure. New calcium dust extinction models are currently being developed (Chu et al.\ in prep.) for future studies of the Ca edge with the next generation of high-resolution X-ray spectrometers.

\section{Summary} \label{sec:summary}

We presented a joint analysis of the Si, S, and Fe K-edge regions in the stacked \chandra/HETG spectrum of \gx, combining $\sim$300~ks of HEG and MEG observations spanning 2001--2023. By fitting the three edges simultaneously, we leveraged the high resolving power at the Si~K edge to constrain the chemistry and crystallinity of silicates, while using the S and Fe K-edge optical depths to refine the contribution from non-silicate compounds.

Our main results are as follows:
\begin{itemize}
\item The Si~K-edge XAFS strongly favors olivine-type silicates over pyroxenes. Amorphous olivine is the dominant dust component, accounting for $\gtrsim60\%$ of the total dust column density in all statistically selected models.
\item The inferred iron budget is dominated by Fe incorporated into silicates ($\sim 74\%$), with an additional contribution from Fe-bearing sulfides and metallic iron ($\sim 8\%$ and $\sim 18\%$, respectively). Given the limited HETG resolution at the Fe~K edge, metallic and gaseous Fe remain partially degenerate, but depletion arguments support a dust-dominated iron budget along this highly-absorbed sightline.
\item The S~K edge shows clear evidence for low-ionization sulfur, with prominent \Sii\ K$\beta$ and K$\gamma$ features. Our best-fit models favor sulfur-bearing dust in the form of iron sulfides (pyrrhotite/troilite), and we infer a sulfur depletion of $\sim35\%$ into the condensed phase.
\item The stacked HEG spectrum also reveals tentative absorption structure associated with low-abundance elements, including the Ar~K-edge region and a marginal detection of the Ca~K edge, highlighting the importance of accurate atomic photoabsorption data at these energies.
\end{itemize}

%% Please use the acknowledgment and contribution environments. This will 
%% be anonomyized when the "anonymous" style option is used. 
\begin{acknowledgments}
We thank the anonymous referee for the careful reading of the manuscript and for the constructive comments, which helped improve the clarity and quality of this work. D.R. is thankful to Efrain Gatuzz for providing and advising on the sulfur photoabsorption cross sections, Lia Corrales for useful discussion on the source and sulfur edge analysis, and Ming-Feng Gu for his help with the FAC calculations. Bruce Draine is thanked for useful discussion on the fraction of Fe in ferromagnetic form. D.R. also thanks Dave Huenemoerder for his insightful suggestions on the reduction of the HETG spectra. During this project, D.R. received support from the Margaret Burbidge Fellowship supported by the Brinson Foundation, as well as NASA through the Smithsonian Astrophysical Observatory (SAO) contract SV3-73016 to MIT for support of the Chandra X-ray Center (CXC) and science instruments.

\end{acknowledgments}

%\begin{contribution}
%%This section gives authors the space to recognize author contributions. The text inside this environment is NOT counted towards the total word quanta. At a minimum, manuscripts are expected to include this text:

%All authors contributed equally to the Terra Mater collaboration.

%% But authors are expected to provide more specific details, e.g. 
%%
%%SC was responsible for writing and submitting the manuscript.
%%WWM came up with the initial research concept and edited the manuscript.
%%OTS obtained the funding and edited the manuscript.
%%EBF provided the formal analysis and validation. He also edited the manuscript.
%%GEH Supervised the undergraduates, wrote the software and administers the project github and Zenodo repositories.
%%
%% Authors can use the Contributor Role Taxonomy (CRediT) at
%% for ideas on how write a good statement tailored to their needs.

%\end{contribution}

\facility{Chandra(HETG).}

\software{astropy \citep{Astropy13,Astropy18,Astropy22},  
          CIAO \citep{Fruscione06}
          \spex \citep{Kaastra22}
          }


%\appendix

%% For this sample we use BibTeX plus aasjournalv7.bst to generate the

\bibliography{biblio}{}
\bibliographystyle{aasjournalv7}



\end{document}